\documentclass[%
 aip,
 amsmath,amssymb,
 reprint,%
]{revtex4-1}

\usepackage{graphicx}
\usepackage{dcolumn}
\usepackage{bm}

\usepackage[utf8]{inputenc}
\usepackage[T1]{fontenc}
\usepackage{mathptmx}
\usepackage{etoolbox}
\usepackage{color}
\usepackage{siunitx}

\newcommand{\ex}[1]{\mathrm{e}^{#1}}

\newcommand{\dd}[0]{\mathrm{d}}

\newcommand{\FF}[0]{\boldsymbol{F}}
\newcommand{\EE}[0]{\boldsymbol{E}}
\newcommand{\ee}[0]{\hat{\boldsymbol{e}}}
\newcommand{\rr}[0]{\boldsymbol{r}}

\newcommand{\vv}[0]{\boldsymbol{v}}

\newcommand{\VV}[0]{\boldsymbol{V}}

\newcommand{\kB}[0]{k_{\mathrm{B}}}

\newcommand{\cgb}[0]{\mathbfcal{G}}
\newcommand{\cg}[0]{{\mathcal{G}}}

\usepackage{makecell}
\DeclareMathAlphabet{\mathcal}{OMS}{cmsy}{m}{n}
\DeclareMathAlphabet\mathbfcal{OMS}{cmsy}{b}{n}

\definecolor{darkblue}{rgb}{0,0,0.6}
\definecolor{darkred}{rgb}{0.6,0,0}
\usepackage[colorlinks=true,urlcolor=darkblue,citecolor=darkblue,linkcolor=darkred]{hyperref}

\makeatletter
\def\@email#1#2{%
 \endgroup
 \patchcmd{\titleblock@produce}
  {\frontmatter@RRAPformat}
  {\frontmatter@RRAPformat{\produce@RRAP{*#1\href{mailto:#2}{#2}}}\frontmatter@RRAPformat}
  {}{}
}%
\makeatother
\begin{document}


\title[Chemotactic particles as strong electrolytes: Debye-H\"uckel approximation and effective mobility law]{Chemotactic particles as strong electrolytes:\\ Debye-H\"uckel approximation and effective mobility law}
\author{Pierre Illien}
\affiliation{Sorbonne Universit\'e, CNRS, Laboratoire PHENIX (Physico-Chimie des \'Electrolytes et Nanosyst\`emes Interfaciaux), 4 Place Jussieu, 75005 Paris, France}
\email{pierre.illien@sorbonne-universite.fr}

\author{Ramin Golestanian}

\affiliation{Department of Living Matter Physics, Max Planck Institute for Dynamics and Self-Organization, D-37077 Göttingen, Germany}
\affiliation{Rudolf Peierls Centre for Theoretical Physics, University of Oxford, OX1 3PU Oxford, United Kingdom}

\date{\today}

\begin{abstract}
We consider a binary mixture of chemically active particles, that produce or consume  solute molecules, and that interact with each other through the long-range concentrations fields they generate. We analytically calculate the effective phoretic mobility of these particles when the mixture is submitted to a constant, external concentration gradient, at leading order in the overall concentration. Relying on a analogy with the modeling of strong electrolytes, we show that the effective phoretic mobility decays with the square-root of the concentration: our result is therefore a nonequilibrium counterpart to the celebrated Kohlrausch and Debye-H\"uckel-Onsager conductivity laws for electrolytes, which are extended here to particles with long-range nonreciprocal  interactions. The  effective mobility law we derive reveals the existence of a regime of maximal mobility, and could find applications in the description  of nanoscale transport phenomena in living cells.
\end{abstract}

\maketitle

\section{Introduction} 

`Chemically active' particles typically refer to  micro- or nano-objects that produce or consume smaller solute molecules, and that interact through the resulting concentrations fields. They are found at different length scales, from active colloids \cite{Zottl2016a, Bechinger2016} down to biological molecules, and especially enzymes, which became paradigmatic examples of nanomotors in the physical and chemical literature \cite{Agudo-Canalejo2018a,{Feng2020,Ghosh2021}}. From a theoretical perspective, chemically active particles have initially been studied under the angle of diffusiophoresis, i.e. through their individual response to external or self-generated drive \cite{Anderson1982a,Anderson1989, {Golestanian2005}, {Golestanian2007},{Ruckner2007a}, {Banigan2016},{Popescu2016},Illien2017,{Ibrahim2017},{Kapral2013},{Oshanin2017,Robertson2020},{Moran2016}}. More recently, the interactions between such particles, and the resulting collective properties, have been investigated. Numerical studies of mixtures of chemically active particles of different species revealed their propensity to phase separate \cite{Saha2014,{Pohl2014}, Agudo-Canalejo2019}, their ability to form complex supramolecular structures that affect their diffusivity \cite{Soto2014,Soto2015,Benois2023},  and quickly became an example of a system with nonreciprocal interactions. Subsequent experimental realizations confirmed the richness and complexity of these collective behaviors \cite{Yu2018,Meredith2020}.

From a theoretical perspective, explicit analytical results on collective effects in such nonequilibrium mixtures are still scarce, due do the technical difficulties raised in the modeling of chemically active particles. Indeed, their pair interactions are generally long-ranged, and should typically be modeled as non-reciprocal \cite{Soto2014,Agudo-Canalejo2019}. In particular, the transport coefficients of chemotactic particles, which characterize their response to external drive, have only been studied elusively, in spite of their importance to understand situations of biological interest, in which spatial heterogeneities are predominant and govern spatial organization \cite{Sear2019}.

\begin{figure}[b]
\begin{center}
\includegraphics[width=\columnwidth]{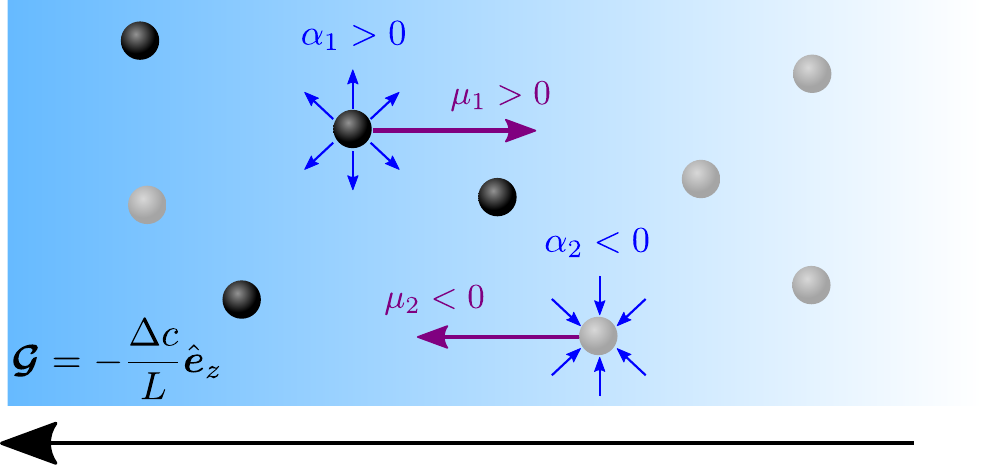}
 \caption{A binary mixture of chemotactic particles placed in a gradient of solute concentrations. As an example, we represent here the situation where particles of species $1$ (resp. $2$) represented in black (resp. grey) tend to produce (resp. consume) solute molecules, and migrate towards low (resp. high) solute concentrations.}
\label{fig_model}
\end{center}
\end{figure}

Here, we rely on the analogy between the physics of chemically active particles and that of strong electrolytes, as has been elucidated in a number of different contexts\cite{Golestanian2012,Grosberg2020}, and we derive an explicit expression for the effective mobility of chemotactic particles in a mixture made of two different species, with long-range, unscreened interactions. In particular, we show that the hydrodynamic and chemotactic interactions between the particles result in a square-root dependence of the effective mobility on the overall volume fraction of particles. Our result is therefore an analogue of the celebrated Kohlrausch \cite{Robinson2002,Bockris,Harned1943} or Debye-H\"uckel-Onsager \cite{Onsager1927, Onsager1932} conductivity laws for electrolytes, which are extended here to an intrinsically nonequilibrium system, with nonreciprocal interactions. We show that the approach to phase separation is associated with a divergence of the effective mobility, and that the parameters that tune the activity of the particles can optimize the mobility of the particles.

\section{Model}

We consider a binary mixture of  active, chemotactic particles (Fig. \ref{fig_model}). There are $N_i$ particles of type $i\in\{A,B\}$, and we denote by $N=N_A+N_B$ their total number. Their positions are denoted by $\rr^i_n$, where $n \in \{1,\dots, N_i\}$, and we denote by $\rr^N = \{\rr_1^A, \dots, \rr_{N_A}^A,\rr_1^B, \dots, \rr_{N_B}^B\}$ the overall configuration.  The overall density of particles of type $i$ is denoted by $n_i = N_i/V$, where $V$ is the volume of the system. We assume that each of the particles can be a source or a sink of smaller solute molecules, whose local concentration is denoted by $c(\rr,t)$. The activity $\alpha_i$, which is homogeneous to an inverse time, characterises the rate at which the particles of type $i$ produce (if $\alpha_i>0$) or consume (if $\alpha_i<0$) solute molecules. Throughout the paper, we will assume that the overall concentration of solute molecules is constant, i.e. that the condition $n_A\alpha_A+n_B\alpha_B=0$ holds. This implies that $\alpha_A\alpha_B<0$, i.e. that one species is a producer of solute molecules while the other one is a consumer.

We assume that each particle interacts with the small solute molecules (for instance because of short-range repulsion interactions), and therefore that diffusiophoretic effects occur: any gradient of solute concentration  results in a net drift of the particles. We will assume that particles of species $A$ and $B$ do not respond in the same way to these gradients. Consequently, when coarse-graining the dynamics by integrating out the degrees of freedom corresponding to the solute molecules, the effective interactions between $A$ and $B$ particles should be assumed to be generally non-reciprocal. Namely, in this class of nonequilibrium systems $\VV_{j\to i}(\rr)$ represents the contribution to the drift velocity experienced by a particle of species $i$ originating from a particle of species $j$, when they are separated by a distance $\rr$.

The starting point of our analysis is the $N$-body Smoluchowski equation obeyed by $\mathcal{P}(\rr^N;t)$, namely the probability to observe the system in a given configuration $\rr^N$ at time $t$, which reads, in the absence of external force:
\begin{align}
&    \partial_t \mathcal{P}(\rr^N;t) = \sum_{k=1}^{N} \Big\{D_{\sigma(k)} \nabla_{\rr_k}^2 \mathcal{P} - \nabla_{\rr_k}\cdot[\mathcal{P}\VV(\rr_k) ] \nonumber\\
  &-\nabla_{\rr_k} \cdot \left [\mathcal{P}\sum_{k'\neq k} \VV_{\sigma(k')\to\sigma(k)} (|\rr_k-\rr_{k'}|)  \right] \Big\},
\label{Smolu}
\end{align}
where $\sigma(k)$ denotes the species of particle $k$ (for instance $\sigma(k)=A$ if $1\leq k \leq N_A$ and $\sigma(k)=B$ if $N_A+1\leq k \leq N_A+N_B$), $D_i$ (connected to the mobility $M_i$ via Einstein relation) denotes the bare diffusion coefficient of particles of species $i$. The second term in the sum represents advection by the solvent, whose velocity field is denoted by $\VV$. Finally, the last term in the sum accounts for the net drift velocity experienced by particle $k$.

We first analyze a mean-field approximation of the Smoluchowski equation \eqref{Smolu}. To this end, we define the one-body densities of particles of species $i$ as $\rho_i(\rr^i_1, t) = \int \prod_{\rr_k \neq \rr^i_1}\dd \rr_k \; \mathcal{P}(\rr^N,t)$. At the mean-field level, the density of particles of type $i$ obey the equation (see Appendix \ref{appendix_smolu}):
\begin{align}
\partial_t \rho_i(\rr,t) =D_i  \nabla^2\rho_i -\nabla\cdot [\VV(\rr)\rho_i] - \nabla\cdot[\vv_i^\text{ph}\rho_i],
\label{eq_rho_i}
\end{align}
where the drift velocity $\vv_i^\text{ph}$ contains the effect of all pair interactions between the particles. For arbitrary interaction potentials between the particle, it is in practice impossible to compute this term. However, if we assume that the particles only interact through the inhomogeneities of the solute field $c(\rr,t)$ (i.e. if we neglect any direct interactions such as short-range repulsion, which is acceptable in the low-density limit), the quantity $\vv_i^\text{ph}$ can be interpreted as a diffusiophoretic velocity, and computed using the classical theory by Derjaguin and Anderson \textit{et al.}\cite{Derjaguin1947,Anderson1989, Anderson1982a, Illien2017, Golestanian2019}. Under these approximations, one gets $\vv_i^\text{ph} \simeq - \mu_i \nabla c$, where $\mu_i$ is a phoretic mobility, and characterizes the response of the particle to inhomogeneities in the solute concentration (this typically holds for short-range interactions between the chemotactic particles and the solute molecules).  If $\mu_i >0$ (resp. $\mu_i<0$), then particles of species $i$ are directed towards (resp. away) regions of lower concentrations. Within this framework, the drift velocity then reads $\VV_{j\to i} = -\mu_i\nabla c_j$, where $c_j(\rr)$ is the solute concentration at position $\rr$ due to the sole presence of a particle of species $j$ at the origin.

\section{`Debye-H\"uckel' approximation}

In order to get insight from the analogy with electrostatics, we first study the equilibrium distribution of chemotactic particles, and ignore the effect of solvent advection ($\VV=0$). The densities $\rho_i$ obey the equation $\partial_t \rho_i(\rr,t) =D_i  \nabla^2\rho_i +\nabla \cdot[(\mu_i \nabla c)\rho_i]$. The solute density is the solution of the reaction-diffusion equation: $\partial_t c(\rr,t) = D_s\nabla^2 c(\rr,t)+\sum_{i} \alpha_i \rho_i(\rr,t)$, where $D_s$ is the diffusion coefficient of solute molecules. If these molecules diffuse very fast (for instance if they are very small compared to the particles, i.e. $D_s \gg D_A,D_B$), one can assume that its concentration obeys the stationary equation
\begin{equation}
   - \nabla^2 c(\rr,t) = \frac{1}{D_s}\sum_{i} \alpha_i \rho_i(\rr,t),
    \label{Poisson_equivalent}
\end{equation}
which is analogous to Poisson's equation for electrolytes. The mapping between the two classes of problems (chemotactic particles and strong electrolytes) is detailed in Appendix \ref{analogy}.

It is not possible to solve explicitly for  the solute concentration $c$ and the density of particles of type $\rho_i$, since the equations they obey are coupled nonlinearly. Linear stability analysis of this set of equations was the object of past work \cite{Agudo-Canalejo2019,vincentjaime-epje-2021}, which revealed the conditions under which such a mixture phase separates. In the stationary state, the densities $\rho_i (\rr,t)$ can be expressed in terms of $c$ as $\rho_i (\rr,t)\simeq n_i \exp \left(  -\beta \frac{\mu_i}{M_i} c(\rr,t) \right)$,
with $\beta = (\kB T)^{-1}$. This amounts to assuming that each particle only interacts with the others through the concentration field of solute molecules, which is acceptable in the low-density limit.  Using this expression of $\rho_i$ in the stationary equation for $c$ yields the Helmholtz equation 
\begin{equation}
-\nabla^2 c(\rr,t) =\frac{1}{D_s}\sum_{i} \alpha_i n_i \ex{  -\beta \frac{\mu_i}{M_i} c(\rr,t) }.
\end{equation}
In order to solve for $c$, a possible strategy is to proceed by analogy with electrostatic interactions, and to apply a `Debye-H\"uckel approximation' by linearizing the exponentials in the rhs of this equation, {and neglecting terms of order $c^2$, which is valid in the low-density limit or when the strength of chemical interactions is typically smaller than the thermal energy. Using the fact that the overall quantity of solute is assumed to be constant ($\sum_i \alpha_i n_i=0$), one gets $\nabla^2 c(\rr,t) = \kappa^{2} c(\rr,t)$, where $ \kappa^{-1} = \left(\sum_i \frac{\alpha_i \beta n_i \mu_i}{D_s M_i}\right)^{-1/2}$  is a screening length, i.e. the typical length over which the effect of the activity of a particle can be felt. Note that, depending on the choice of parameters, $\kappa^{-2}$ may be negative: this corresponds to the situation where the mixture of chemotactic particles is unstable \cite{Agudo-Canalejo2019,vincentjaime-epje-2021}.

\section{Response to an external concentration gradient}

We now assume that the chemotactic particles are placed in a constant external gradient of solute concentration, that we denote by $\cgb= -\frac{\Delta c}{L} \ee_z$, where ${\Delta c}$ is the  difference of concentrations imposed at the two ends of the system, and  $L$ is the typical size of the system in the $z$ direction. A particle of species $i$ is then submitted to the external drift velocity $\VV_i^\text{ext} = \mu_i \cgb$.
We denote by $v_i(\cg)$ the stationary velocity of a particle of type $i$ along direction $z$ when one applies a concentration gradient of magnitude $\cg$. The effective mobility of the particle is defined as $\mu_i^\text{eff} = \lim_{\cg\to 0} v_i(\cg)/\cg$ and is affected by the presence of other particles because of two main effects \cite{Fuoss1959}: a `hydrodynamic' effect, that originates from the viscous drag in the suspension; and a `chemotactic' effect, that originates from the interactions mediated by the inhomogeneities in the spatial distribution of solute molecules. Therefore, the  velocity of a particle of $i$ is decomposed as: $\vv_i = \vv_i^0+\vv^\text{hyd}_i+\vv^\text{chem}_i.$ A similar decomposition has been used in the exact solution of non-reciprocal phoretic interactions between two finite-sized colloids\cite{BabakPhysRevLett.124.168003}. In what follows, we present derivation of the two velocity increments $\vv^\text{hyd}_i$ and $\vv^\text{chem}_i$.

\subsection{Hydrodynamic contribution}

From the `Debye-H\"uckel' approximation, we know that each particle will be surrounded by a spherical cloud of `counter-particles', of typical radius $\kappa^{-1}$. Therefore, when a particle moves in a concentration gradient, it also drags a large number of `counter-particles' with itself. The resulting hydrodynamic drag tends to slow down the particle, which undergoes a drag force which will typically be directed against $\VV_i^\text{ext}$, and that is proportional to $\vv^\text{hyd}_i$. In order to estimate the prefactor in front of $\vv^\text{hyd}_i$, we approximate the cloud as a sphere of radius $\kappa^{-1}$, and by analogy with the drag experienced by a solid sphere in the Stokes limit, we estimate the drag force acting on the colloidal particle and its counter-particles as $-6\pi \eta \kappa^{-1} \vv^\text{hyd}_i$. 
At the steady state, this force will be balanced against the drive $(\mu_i/M_i)\cg$. We then estimate the hydrodynamic drift velocity as $\vv^\text{hyd}_i \simeq - \frac{\mu_i}{6\pi \eta M_i\kappa^{-1} } \cg$.

\subsection{Chemotactic contribution}

When the considered particle moves (either because of the applied external gradient, or because of thermal fluctuations), it is displaced from the center of its `cloud' of counter-particles. Indeed, the latter need some time to respond to the displacement of the central particle and to re-arrange. The electrostatic analogue to this effect is the relaxation of a cloud of counterions to its spherical equilibrium shape, after it is perturbed by an external electric field. The particle will tend to be brought back towards the center of the cloud, and will undergo a force  directed against $\VV_i^\text{ext}$.  For a particle located at $\rr=0$, the resulting velocity increment $\vv^\text{chem}_i$ reads $\vv^\text{chem}_i=-\mu_i\left.\nabla c_i'\right|_{\rr=0}$, where we denote by $c'_i=c_i-c_i^0$ the perturbation induced on the equilibrium solute distribution by the external field.

The perturbation to the solute distribution $c'_i(\rr,t)$ is computed by treating the problem at the two-particle level, i.e. beyond the  mean-field approximation presented above. To this end, we define the two-particle distribution $f_{ji}(\rr_1,\rr_2;t)$, namely the probability to find a particle of type $j$ at $\rr_1$ and a particle of type $i$ at $\rr_2$ at time $t$. It is related to the $N$-body distribution through $f_{ji}(\rr^j_1,\rr^i_2;t) = \int \prod_{\rr_k \notin \{ \rr^i_1,\rr^j_2\} }\dd \rr_k \; \mathcal{P}(\rr^N,t)$. Performing the corresponding integration on Eq. \eqref{Smolu}, and neglecting the three-body distributions as a first approximation, yields the following equation for the two-body distributions:
\begin{align}
\label{twobod_noapprox}
&\partial_t f_{ji}(\rr_1,\rr_2;t) = D_j \nabla_{\rr_1}^2 f_{ji}+D_i \nabla_{\rr_2}^2 f_{ij} \\
& -  \nabla_{\rr_1}\cdot f_{ij}\,{\mu_j} (\cgb- \nabla_{\rr_1} c'_j(0)-\nabla_{\rr_1}c_i(\rr_1,\rr_2))  \nonumber\\
& -  \nabla_{\rr_2}\cdot f_{ji}\,{\mu_i} (\cgb- \nabla_{\rr_2} c'_i(0)-\nabla_{\rr_2}c_j(\rr_1,\rr_2)), \nonumber
\end{align}
where we neglected the effect of solvent flow ($\VV(\rr) \simeq 0$), as we assume that the effect of hydrodynamic entrainment is correctly captured by the study of the hydrodynamic contribution detailed above.

\begin{figure*} 
\includegraphics[width=0.32\textwidth]{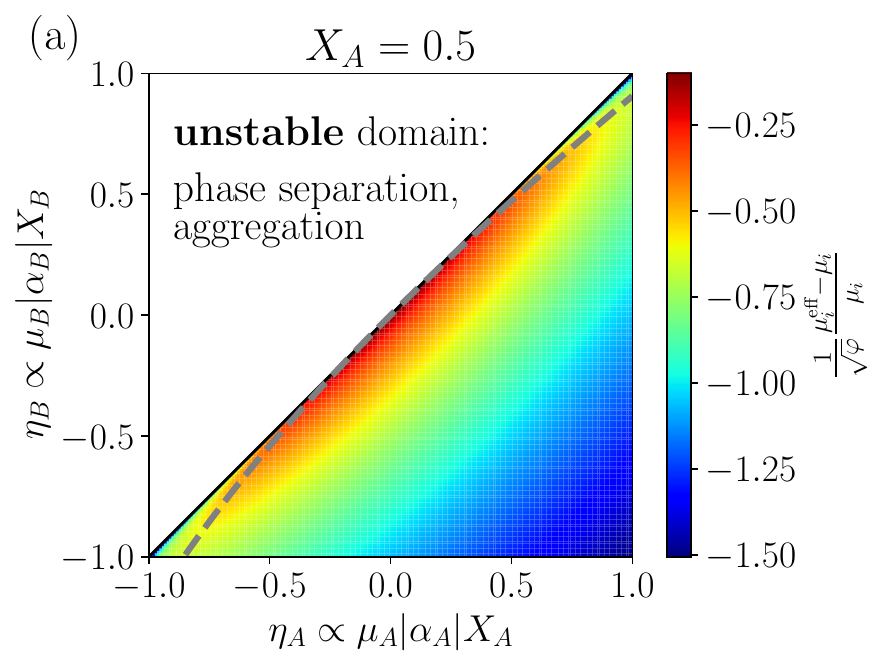}
\includegraphics[width=0.305\textwidth]{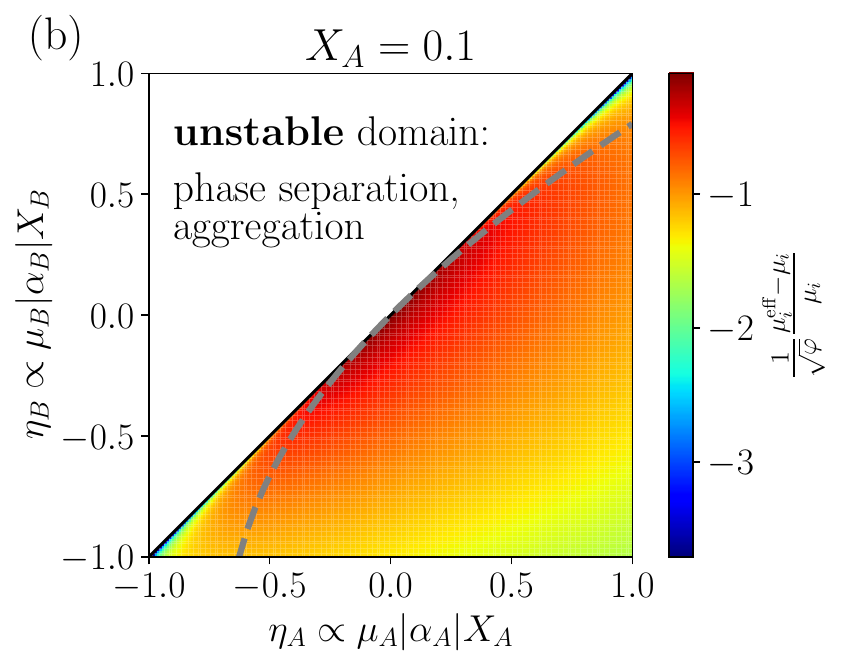}
\includegraphics[width=0.32\textwidth]{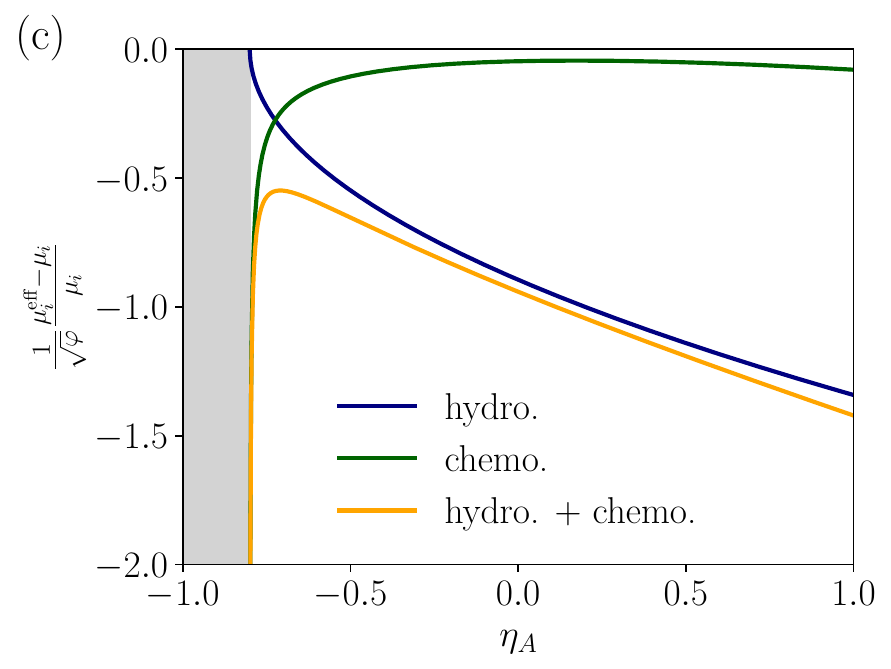}
\caption{Effective mobility $\mu_i^\text{eff}$ of the chemotactic particles rescaled by its bare value $\mu_i$ and by square root of concentration, as a function of the dimensionless variables $\eta_A$ and $\eta_B$, for different values of the relative fraction of $A$ and $B$ particles: (a) $X_A=0.5$; (b) $X_A=0.1$. The dashed gray lines represents the regime of maximum mobility.  (c) Same quantity represented as a function of $\eta_A$, for a fixed value of $\eta_B$ ($\eta_B=-0.8)$ and for $X_A=0.5$. }
\label{fig_effective_mob}
\end{figure*}

Assuming that the perturbation induced by the external drive is small, we write the distribution functions as $f_{ji}(\rr_1,\rr_{21}) = f_{ji}^0(r) + f'_{ji}(\rr_1,\rr_{21}) $,
where we denote $r=|\rr_2-\rr_1|$, and where $f'_{ji} \ll f_{ji}^0$. The equilibrium solute distribution is determined within our `Debye-H\"uckel' approximation, i.e. in the limit or small concentration or of sufficiently weak interactions, and the equilibrium pair distribution is given by 
\begin{equation}
    f_{ji}^0 (r) \simeq n_i n_j \left(1-  \frac{\mu_i}{\kB T M_i}\frac{\alpha_j}{4\pi D_s} \frac{\ex{- \kappa r}}{r}\right).
    \label{f0}
\end{equation}
Importantly, note that non-reciprocal interactions translate into the pair distribution functions, with $f_{AB}$ being generally different from $ f_{BA}$.  Next, under the assumption that the applied concentration field is weak, we perform a perturbative expansion and only retain terms at linear order in $c'_j$ and $f'_{ji}$.  Taking the stationary limit of Eq. \eqref{twobod_noapprox}, one gets
 \begin{align}
& \mu_i \cgb\cdot \nabla_{\rr_2}  f^0_{ji} 
 +\mu_j  \cgb\cdot \nabla_{\rr_1}  f^0_{ij}  \nonumber\\
 &-\mu_i n_i n_j \nabla_{\rr_2} ^2 c'_j - \mu_j n_i n_j \nabla_{\rr_1} ^2 c'_i\nonumber \\
&-M_i \kB T \nabla_{\rr_2}^2 f'_{ji}-M_j \kB T \nabla_{\rr_1}^2 f'_{ij} = 0
\end{align}
Finally, $f'_{ij}$ is eliminated by using the analogous to Poisson's equation written at linear order in the perturbation: $-\nabla^2 c'_j = \frac{1}{D_s} \sum_i \frac{\alpha_i}{n_j} f'_{ij}$. One gets a set of coupled differential equations for the perturbations $c'_i$ and $c'_j$, which are solved to yield (see Appendix \ref{appendix_chemo}):
\begin{equation}
 (\nabla c'_i)_{\rr=0} = \frac{\kappa}{12\pi D_s \kB T}\frac{\mu_A^2 \alpha_B/M_A-\mu_B^2\alpha_A/M_B}{\mu_A-\mu_B}\frac{\mathcal{Q}}{1+\sqrt{\mathcal{Q}}}\cg
 \label{contrib_chem}
\end{equation}
for $i\in \{ A,B\}$, with the dimensionless number 
\begin{equation}
    \mathcal{Q} = \frac{M_A M_B}{M_A + M_B}  \frac{\mu_A-\mu_B}{\mu_A M_B - \mu_B M_A}.
\end{equation} 
The resulting velocity increment of a particle of type $i$ along direction $\ee_z$ is simply given by $v_i^\text{chem} = -\mu_i  (\nabla c'_i)_{\rr=0} $. 
We emphasize that $\mathcal{Q}$ may become negative, which would result in the chemotactic contribution to mobility having a nonzero imaginary part. Interestingly, and although we will not investigate the corresponding range of parameters here, this could be the signature of emerging oscillatory (underdamped) dynamics, in a similar vein to oscillatory instabilities that have been reported elsewhere\cite{OuazanReboul2023}.

\section{Effective mobility}

We now put together the hydrodynamic and chemotactic contributions and give the expression of the effective mobility. We focus on the situation where the $A$ and $B$ particles have the same size and therefore the same mobilities $M_0$ and diffusion coefficients $D_0$ (this implies $\mathcal{Q}=1/2$). The effective mobility can be rewritten in terms of the  variables $\eta_j \equiv \frac{3|\alpha_j| \mu_j X_j}{4\pi D_s D_0 a}$,
where $X_j = n_j/(n_1+n_2)$ is the fraction of particles of species $j$ and $a$ is the radius of a particle. It reads:
\begin{align}
&\frac{\mu_i^\mathrm{eff}}{\mu_i} =1-\sqrt{\varphi} \left[\sqrt{\eta_A-\eta_B} + \dfrac{\sqrt{2}-1}{9\sqrt{2}} \dfrac{\eta_A^2 X_A + \eta_B^2 X_B}{X_AX_B \sqrt{\eta_A-\eta_B}}   \right] ,
\label{general_meff_rewritten}
\end{align}
where we assumed, with no loss of generality, that  $\alpha_A >0$, $\alpha_B<0$, and where $\varphi \equiv \frac{4}{3}\pi a^3 (n_A+n_B)$ is the overall volume fraction of chemotactic particles. Several comments follow: (i) The expression of the effective mobility given by Eq. \eqref{general_meff_rewritten} constitutes the central result of this article. The result is valid at low-densities, namely at order $\mathcal{O}(\sqrt{\varphi})$ when $\varphi\ll 1$, and in the limit where the interactions between the particles are sufficiently small as compared with the thermal energy. With the choice of $\alpha_A\alpha_B<0$, and in the absence of external field, in the domain $\eta_A>\eta_B$, the suspension remains homogeneous, and the particles form small `molecules', that self-propel or rotate depending on their symmetry properties \cite{Soto2014,Soto2015}. On the contrary, in the domain $\eta_A<\eta_B$, the suspension is unstable and separates into dilute regions and dense, large clusters\cite{Agudo-Canalejo2019}.  (ii) Note that, in the limit we consider here, the effective mobility rescaled by its value at infinite dilution $\mu_i^\text{eff}/\mu_i$ is identical for both species. This identity is not expected to be generally true, and would  not hold for higher concentrations, or for stronger chemical gradients, beyond linear response. (iii) As expected naively, the effective mobility decreases when the volume fraction increases: collective effects generally hinder the transport of chemotactic particles under the effect of an external gradient. However, the mobility decays as the square root of volume fraction. This effect, which is a consequence of the square root dependence of the screening length $\kappa^{-1}$, is reminiscent from the celebrated conductivity laws in electrolytes, first intuited and observed experimentally by Kohlrausch\cite{Robinson2002,Bockris,Harned1943}, and later derived theoretically from microscopic considerations by Debye and H\"uckel -- a theory subsequently refined by Onsager\cite{Onsager1927, Onsager1932}. We believe that the present result should be seen as an extension of this central result to a more general setting, in which interactions are nonreciprocal and intrinsically nonequilibrium.

Our analytical result \eqref{general_meff_rewritten} takes a simpler form in the case of a `symmetric mixture' where $n_A=n_B=n$, i.e. $X_A=0.5$ (which imposes $\alpha_A = -\alpha_B =\alpha$, with $\alpha>0$). In this situation, one can define the length $\ell =\frac{3 \mu_A\alpha}{8\pi D_s D_0}$, which is the typical distance at which the interactions between chemotactic particles becomes comparable to the thermal energy $\kB T$. In this respect, it can be seen as the analogous to the `Bjerrum length' in electrostatics. With this definition, the effective mobility has a more explicit expression, and reads
\begin{align}
&\frac{\mu_i^\mathrm{eff}}{\mu_i} =1-\sqrt{\varphi} \sqrt{\frac{\ell}{a}}\left[\sqrt{1-\mathcal{M}} + \dfrac{\sqrt{2}-1}{9\sqrt{2}} \frac{\ell}{a}  \frac{1+\mathcal{M}^2}{\sqrt{1-\mathcal{M}}} \right] ,
\label{general_meff_rewritten_sym}
\end{align}
where we define the ratio between the two mobilities  $\mathcal{M} = \mu_B/\mu_A$. In the limit $\mathcal{M}\to 1$ (i.e. when both species have comparable mobilities, but opposite activities), we observe that the hydrodynamic contribution to the mobility vanishes -- this is expected, since all the particles respond similarly to the imposed external gradient, and the particles do not have to `drag' a cloud of counter-particles anymore. On the contrary, the chemotactic contribution increases in magnitude, and significantly decrease the effective mobility: the formation of large and cohesive clusters at the approach of the phase separation threshold tend to arrest particles and hinder their response to external perturbations. In the opposite limit $\mathcal{M}\to -\infty$ (i.e., for a fixed value of $\mu_A$, when the $B$ particles have a very strong negative response), both hydrodynamic and chemotactic contributions penalise the effective mobility: this is interpreted by the formation of very cohesive  `molecules' made of $A$ and $B$ `atoms', which are less sensitive to external gradients. The divergence of the mobility ratio in this limit, which is singular and unphysical, could be regularized by considering higher-order terms in our small-$\varphi$ expansion. Finally, since the effective mobility appears to be decreasing function of $\mathcal{M}$ in both limits of $\mathcal{M} \to -\infty$ and $\mathcal{M}\to 1$, one expects the existence of a regime of optimal mobility.

We plot on Fig. \ref{fig_effective_mob}  the effective mobility rescaled by its bare value and by square root of concentration $(\mu_i^\text{eff}-\mu_i)/(\sqrt{\varphi} \mu_i)$, as given in Eq. \eqref{general_meff_rewritten}, as a function of the dimensionless variables $\eta_A$ and $\eta_B$. For $X_A=0.5$, the qualitative behavior that we read from Eq. \eqref{general_meff_rewritten_sym} is confirmed by the plot on Fig. \ref{fig_effective_mob}(a), and also holds for a more general asymmetric mixture [Fig. \ref{fig_effective_mob}(b)]: between the regimes two regimes of  $\eta_A \simeq \eta_B$ and $\eta_A \ll \eta_B$, in which the effective mobility is significantly reduced, there exists an regime of optimal mobility. The relative contributions of the hydrodynamic and chemotactic effects are shown on Fig. \ref{fig_effective_mob} (c), for a fixed value of $\eta_B$.

\section{Conclusion}

We highlight the fact that the mobility law we derived through an analogy with strong electrolytes is  general and applies to chemotactic particles at different lengthscales: from molecular machines, such as enzymes (a few nm), to active colloids (a few microns), the main hypothesis being that solute molecules diffuse fast enough and equilibrate quickly. Our main result could be of particular interest in biological situations, where phoretic mechanisms and chemical interactions are known to govern spatial organization and metabolic pathways \cite{Agudo-Canalejo2018a,Zhao2018a,Burkart2022,Ramm2021,OuazanReboul2023}. From a technical point of view, we underline the formal analogy between the present system and a simple electrolyte. Relying on classical electrochemistry works, this opens the way to many analytical developments. For instance, it would be particularly interesting to go beyond the dilute limit and to incorporate the effect of short-range repulsion between particles, that will predominate in concentrated suspensions. We could rely on the large body of literature that have been devoted to this topic in the context of electrolytic solutions \cite{Bernard1992,Bernard1992a,Chandra1999,Dufreche2002a,Dufreche2005,Zorkot2018,Avni2022,Bernard2023}.

\begin{acknowledgments}
P. I. acknowledges fruitful discussions with Olivier Bernard, Marie Jardat, and Benjamin Rotenberg.
\end{acknowledgments}

\appendix

\section{Mean-field treatment of the $N$-body Smoluchowski equation}
\label{appendix_smolu}

We start from the $N$-body Smoluchowski equation satisfied by the distribution $ \mathcal{P}(\rr^N;t)$ [Eq. \eqref{Smolu} in the main text]:
\begin{align}
    &\partial_t \mathcal{P}(\rr^N;t) = \sum_{k=1}^{N} \Bigg\{D_{\sigma(k)} \nabla_{\rr_k}^2 \mathcal{P} - \nabla_{\rr_k}\cdot[\mathcal{P}\VV(\rr_k) ] \nonumber\\
&-\nabla_{\rr_k} \cdot \left [\mathcal{P}\sum_{k'\neq k} \VV_{\sigma(k')\to\sigma(k)} (|\rr_k-\rr_{k'}|)  \right] \Bigg\}.
\label{Smolu_supp}
\end{align}
With no loss of generality, we compute the one-body density associated to the $A$ particles: $\rho_A(\rr^A_1, t) = \int \prod_{\rr_k \neq \rr^A_1}\dd \rr_k \; \mathcal{P}(\rr^N,t)$, which reads:
\begin{align}
&\partial_t \rho_A(\rr,t) = D_A\nabla^2 \rho_A - \nabla \cdot[\VV(\rr) \rho_A]
\nonumber\\
&-(N_A-1) \nabla \cdot \int \dd \rr' \; f_{AA}(\rr,\rr',t) \VV_{A\to A}(|\rr-\rr'|) \nonumber\\
&-N_B \nabla \cdot \int \dd \rr' \; f_{AB}(\rr,\rr',t) \VV_{B\to A}(|\rr-\rr'|),
\end{align}
where the two-body distributions are defined in the main text as:
\begin{equation}
f_{ji}(\rr^j_1,\rr^i_2;t) = \int \prod_{\rr_k \notin \{ \rr^i_1,\rr^j_2\} }\dd \rr_k \; \mathcal{P}(\rr^N,t).
\end{equation}

The non-reciprocal drift velocities are here assumed to originate from phoretic interactions\cite{Soto2014}. We assume that the particles primarily interact through the production and consumption of solute molecules, which yield an inhomogeneous distribution, and the resulting phoretic transport. The drift velocity can be determined following the classical approach by Derjaguin and Anderson \cite{Derjaguin1947,Anderson1989, Anderson1982a,Golestanian2019}. Denoting by $c(\rr,t)$ the solute concentration at position $\rr$ and at time $t$, the phoretic velocity can be written as $\vv_A^\text{ph} \simeq -\mu_A \nabla c$. Within this approach, the mobility $\mu_A$ reads\cite{Golestanian2019}
\begin{equation}
    \mu_A = \frac{\kB T}{\eta}\underbrace{\int_0^\infty \dd z \;z( 1-\ex{-\psi_{As}(z)/\kB T})}_{\equiv \lambda_A^2},
\end{equation}
with $\psi_{As}$ the interaction potential between solute molecules and the surface of the particles of species $A$, and where $\lambda_A$ is usually called the `Derjaguin length'. We emphasize that the `mobilities' $\mu_i$ (that have dimension L\textsuperscript{5}T\textsuperscript{-1}) should not be mistaken with the mobilities $M_i$ (with dimension 1/(MT\textsuperscript{-1})) that are usually defined as the ratio between velocity and force within linear response.

\section{Analogy with electrostatics}
\label{analogy}

In this Appendix, we highlight the analogy between the system under study and a binary electrolyte. Consider a binary electrolyte, made of $N_+$ cations of charge $e_+=z_+ e$ and $N_-$ anions of charge $e_-=z_- e$.
The concentration of the cations and anions are denoted by $n_+=N_+/V$ and $n_-=N_-/V$, respectively, where $V$ is the volume of the system. The electroneutrality condition then reads $n_+ e_+ + n_- e_-=0$. The electrostatic potential in the solution obeys the Poisson equation $-\nabla^2 \varphi(\rr) = \rho(\rr)/\varepsilon$, where $\rho(\rr)$ is the charge density at position $\rr$ in the solution. Denoting by $\rr_i^+$ (resp. $\rr_i^-$), with $n=1,\dots,N=N_++N_-$  the position of the ions. The density of ions of species $i=\pm$ is defined as $\rho_i(\rr)=\sum_{n=1}^{N_i} \delta(\rr-\rr_n^i)$. With these notations, the Poisson equation is rewritten
\begin{equation}
\label{eq_Poisson}
- \nabla^2 \varphi(\rr) =\frac{1}{\varepsilon} \sum_i e_i \rho_i(\rr).
\end{equation}
This is analogous to the equation satisfied by the concentration field $c$ in the main text [Eq. \eqref{Poisson_equivalent}].

We write the force experienced by an ion of species $i$ located at position $\rr$: $\FF_i(\rr)=e_i \EE(\rr)=-e_i \nabla \varphi(\rr)$. This is analogous to the pseudo-force introduced in the main text. However, for chemotactic particles, activity and response to field gradients are not controlled by a unique parameter (which is the charge $e_i$ in the electrostatic analogy). A consequence is that chemotactic particles interact in a non-reciprocal way: the interaction exerted by a particle of species $i$ on a particle of species $j$ is generally not the opposite of that exerted by $j$ on $i$. This is obviously not the case for ions. We summarize the analogy between electrostatics and chemotactic particles in Table~\ref{table_analogy}.

\begin{table*}
  \centering 
  \begin{tabular}{l|l}
\hline \hline
 \textbf{chemotactic particles}  & \textbf{electrolytes}   \\
\hline \hline
$D_s$: diffusion coefficient of solute molecules & $\varepsilon$: dielectric permittivity \\  \hline
$\alpha_i$: chemical activity   & $e_i$: charge (source in Poisson's equation)    \\ \hline
$\mu_i$: response to a chemical gradient & $e_i$: charge (response to an electric field) \\ \hline
 $c_i \propto \alpha_i$: solute concentration due a particle of species  $i$  & $\varphi_i \propto e_i$: electrostatic potential due to the presence of an ion $i$      \\  \hline
 $\cgb=-\nabla c$: applied `chemical' field  &  $\EE=-\nabla \varphi$: applied electric field  \\  \hline
\makecell[cl]{ $\VV_{j \to i}  = -\mu_i \nabla c_j \propto \mu_i \alpha_j$: velocity of a particle \\   of species $i$ due to the presence of a particle of species $j$} & \makecell[cl]{ $\FF_{j \to i}  = -e_i \nabla \varphi_j \propto e_i e_j$: force undergone by an ion \\   of species $i$ due to the presence of an ion of species $j$}\\
\hline
\makecell[cl]{ $ \kappa^{-1} = \left(\sum_i \frac{\alpha_i \beta n_i \mu_i}{D_s M_i}\right)^{-1/2}$: screening length of \\chemical interactions }&  {$ \kappa_\text{D}^{-1} = \left(\sum_i \frac{n_ie_i^2}{\kB T \varepsilon}\right)^{-1/2}$: Debye screening length }\\
\hline\hline
\end{tabular}
  \caption{Analogy between the description of chemotactic particles and strong electrolytes.}
  \label{table_analogy}
\end{table*}

\section{Perturbative resolution of the two-body dynamics}
\label{appendix_chemo}

The starting point of our analysis is Eq. \eqref{twobod_noapprox} in the main text, that we consider in the stationary limit:
\begin{align}
&0= D_j \nabla_{\rr_1}^2 f_{ji}+D_i \nabla_{\rr_2}^2 f_{ij} \nonumber\\
& -  \nabla_{\rr_1}\cdot f_{ij} \;\mu_j(\cgb- \nabla_{\rr_1} c'_j(0)-\nabla_{\rr_1}c_i(\rr_1,\rr_2))  \nonumber\\
& -  \nabla_{\rr_2}\cdot f_{ji} \;\mu_i(\cgb- \nabla_{\rr_2} c'_i(0)-\nabla_{\rr_2}c_j(\rr_1,\rr_2)).
\end{align}
We expand this equation at linear order in the perturbations $c'_j$ and $f'_{ji}$. 
It is clear from the expression of the equilibrium distribution $f_0^{ij}$:
\begin{equation}
f_{ji}^0 (r) \simeq n_i n_j \left(1-  \frac{\mu_i}{\kB T M_i}\frac{\alpha_j}{4\pi D_s} \frac{\ex{- \kappa r}}{r}\right),
\label{f0_supp}
\end{equation}
that $f_{ji}-n_i n_j$ is of order $\mu_i\alpha_j/M_i$ so, in the expression $ (\mu_i/M_i) f_{ji} \nabla_{\rr_2}  c_j$, $ f_{ji}$ is replaced by $n_i n_j$. We use Eq. \eqref{f0_supp} to eliminate the terms of order 0, and we get, at linear order in the perturbation\cite{Harned1943}:
\begin{align}
& \mu_i \cgb\cdot \nabla_{\rr_2}  f^0_{ji} 
 +\mu_j \cgb\cdot \nabla_{\rr_1}  f^0_{ij}  -\mu_i n_i n_j \nabla_{\rr_2} ^2 c'_j - \mu_j n_i n_j \nabla_{\rr_1} ^2 c'_i\nonumber \\
&-M_i \kB T \nabla_{\rr_2}^2 f'_{ji}-M_j \kB T \nabla_{\rr_1}^2 f'_{ij} = 0,
\end{align}
where we used the assumption that the external gradient is constant ($\nabla \cdot \cgb =0$). This equation is now used to compute the chemotactic or `relaxation' effect. Since the external gradient imposed to the solution is in the $z$ direction, we write $\cgb=\cg \ee_z$. We rewrite all the quantities in terms of the distance between the two particles $\rr=\rr_2-\rr_1$, and replace the gradient operators as $\nabla_{\rr_2}=\nabla_{\rr}$ and $\nabla_{\rr_1}=-\nabla_{\rr}$  (in what follows, we drop the index $\rr$ for clarity). We use the following symmetry relations: 
\begin{align}
c'_i(\rr) &= -c'_i(-\rr),\\
f'_{ij}(\rr) &= -f'_{ji}(-\rr).
\end{align}
We keep in mind that generally $f_{ij}^0(\rr) \neq f_{ji}^0(\rr) $. 
We ignore the dependence on the solvent velocity field $\VV$, since the effect of hydrodynamic is considered separately. We obtain:
\begin{align}
&\mu_i \cg \frac{\partial}{\partial z} f^0_{ji} -\mu_j \cg  \frac{\partial}{\partial z} f^0_{ij} -\mu_i  n_i n_j \nabla^2 c'_j+\mu_j  n_i n_j \nabla^2 c'_i  \nonumber\\
&-(M_i+M_j)\kB T \nabla^2 f'_{ji}=0.
\label{cont2}
\end{align}
The derivative of $f^0_{ji}$ is computed using Eq. \eqref{f0}:
\begin{equation}
\label{partial_z_f0}
\frac{\partial}{\partial z} f^0_{ji} = -n_i n_j\frac{\mu_j}{\kB T M_j} \frac{\alpha_i}{4\pi D_s}\frac{\partial}{\partial z} \left( \frac{\ex{- \kappa r}}{r}\right).
\end{equation}
Then, the perturbations $f'_{ji}$ are eliminated and replaced by $c'_j$ using the Poisson equation written at linear order in the perturbation:
\begin{equation}
-\nabla^2 c'_j = \frac{1}{D_s} \sum_i \frac{\alpha_i}{n_j} f'_{ji}.
\end{equation}
Multiplying Eq. \eqref{cont2} by $\alpha_i/n_j$, dividing by $(M_i+M_j)$ and summing over $i$ yields
\begin{align}
&\nabla^4 c'_j-\frac{1}{\kB T D_s}\sum_i \frac{\mu_i n_i \alpha_i}{M_i+M_j}\nabla^2 c'_j +\frac{1}{\kB T D_s}\sum_i \frac{\mu_j n_i \alpha_i}{M_i+M_j} \nabla^2 c'_i \nonumber\\
&= -\frac{1}{\kB T D_s} \sum_i \frac{\alpha_i}{n_j(M_i + M_j)} \left(\mu_i \cg \frac{\partial f^0_{ji}}{\partial z}-\mu_j \cg \frac{\partial f^0_{ij}}{\partial z}\right).
\label{ODE_4th}
\end{align}
The rhs can be made explicit by using Eq. \eqref{partial_z_f0}. This yields a closed set of linear equations for the perturbations to the perturbations $c'_i$:
\begin{align}
&\nabla^4 c'_j-\frac{1}{\kB T D_s}\sum_i \frac{\mu_i n_i \alpha_i}{M_i+M_j}\nabla^2 c'_j +\frac{1}{\kB T D_s}\sum_i \frac{\mu_j n_i \alpha_i}{M_i+M_j} \nabla^2 c'_i \nonumber\\
&= \frac{1}{4\pi(\kB T D_s)^2} \sum_i \frac{\alpha_i}{M_i + M_j} \left( \frac{n_i \mu_i^2 \alpha_j}{M_i}-\frac{n_i \mu_j^2 \alpha_i}{M_j}\right)  \nonumber\\
&\times \frac{\partial}{\partial z} \left(   \frac{\ex{-\kappa r}}{r} \right)\cg.
\label{ODE_4th_bis}
\end{align}
For an arbitrary number of species, Eq. \eqref{ODE_4th_bis} can be solved by resorting to a matrix formalism \cite{Onsager1932}. Here, we focus on the  case where there are only 2 species in the mixture. We specify Eq. \eqref{ODE_4th_bis} with $j=A$, and we eliminate $c'_B$. To this end, we multiply Eq. \eqref{ODE_4th_bis} by $n_j \alpha_j$ and sum over $j$. This yields $\sum_j n_j \alpha_j \nabla^4 c'_j=0$. Since $c'_j$ and its Laplacian must remain finite and vanish for $r\to \infty$, this implies $\sum_j n_j \alpha_j \nabla^2 c'_j=0$, i.e., when there are only 2 species: $n_A \alpha_A \nabla^2 c'_A + n_B \alpha_B \nabla^2 c'_B=0$. 
We obtain 
\begin{align}
&\nabla^4 c'_A-\frac{1}{\kB T D_s} \frac{\mu_A n_A \alpha_A + \mu_B n_B \alpha_B}{M_A + M_B} \nabla^2 c'_A\nonumber\\
&=\frac{\cg}{4\pi(\kB T D_s)^2} \frac{\alpha_B n_B}{M_A+M_B}\left(\frac{\mu_B^2 \alpha_A}{M_B} - \frac{\mu_A^2 \alpha_B}{M_A} \right) \frac{\partial}{\partial z} \left(   \frac{\ex{-\kappa r}}{r} \right).
\end{align}
We define:
\begin{equation}
\tilde\kappa^2 = \frac{1}{D_s\kB T} \frac{\mu_A n_A \alpha_A+ \mu_B n_B\alpha_B}{M_A+M_B},
\end{equation}
which has the dimension of an inverse square length. Using the global conservation of the quantity of solute: $n_A \alpha_A+n_B\alpha_B=0$, we find that $\tilde\kappa^2$ is related to the screening length through:
\begin{equation}
\tilde\kappa^2 = \underbrace{\frac{M_A M_B}{M_A + M_B}  \frac{\mu_A-\mu_B}{\mu_AM_B - \mu_B M_A}  }_{\equiv \mathcal{Q}} \kappa^2,
\end{equation}
where $\mathcal{Q}$ is dimensionless. We also introduce:
\begin{equation}
\tilde{\cg} = \frac{1}{4\pi D_s\kB T}\frac{\mu_A^2 \alpha_B/M_A-\mu_B^2\alpha_A/M_B}{\mu_A-\mu_B}\cg.
\end{equation}
We then get 
\begin{equation}
\nabla^4 c'_A - \tilde\kappa^2 \nabla^2 c'_A = \tilde{\cg}  \tilde\kappa^2 \frac{\partial}{\partial z} \left( \frac{\ex{-\kappa r}}{r}\right).
\label{ODE4}
\end{equation}

 The solution of this differential equation with appropriate boundary conditions reads (see Appendix \ref{sol_ODE}), as an expansion in powers of $r$:
\begin{equation}
c'_A = \frac{\tilde{\cg} \tilde\kappa^2}{ \tilde\kappa^2-\kappa^2}\left(-\frac{\kappa- \tilde\kappa}{3}z +  \frac{\kappa^2- \tilde\kappa^2}{8}rz + \mathcal{O}(r^2)\right). 
\label{phi1_exp}
\end{equation}
We are interested in the concentration gradient created by the perturbation of the ionic atmosphere, which reads 
\begin{equation}
 (\nabla c'_A)_{r=0} = \frac{\tilde{\cg}  \kappa}{3} \frac{ \mathcal{Q}}{1+\sqrt{ \mathcal{Q}}}.
\end{equation}

The resulting velocity increment of a particle of type $i$ along direction $\ee_z$ is
\begin{align}
v_i^\text{chem} &=- \mu_i  (\nabla c'_1)_{r=0} \\
&=-\mu_i \frac{\kappa}{3} \frac{1}{4\pi D_s\kB T} \frac{\mu_A^2\alpha_B/M_A - \mu_B^2\alpha_A/M_B}{\mu_A-\mu_B} \frac{\mathcal{Q}}{1+\sqrt{\mathcal{Q}}} \cg,
\label{vrelax}
\end{align}
which is the expression given in the main text [Eq. \eqref{contrib_chem}].

\section{Solution of the differential equation Eq. (\ref{ODE4})}
\label{sol_ODE}

In this Section, we give the solution of the differential equation \eqref{ODE4}. First, using the relation
\begin{align}
\nabla^2 \left( \frac{\ex{- \kappa r}}{r}\right) &= \frac{1}{r^2}\frac{\partial}{\partial r} \left[ r^2 \frac{\partial}{\partial r}\left( \frac{\ex{-\kappa r}}{r}\right)  \right] =\kappa^2\frac{\ex{-\kappa r}}{ r},
\end{align}
it is clear that a particular solution of Eq.  \eqref{ODE4} is
\begin{equation}
c'_A = \frac{\tilde{\cg}\tilde\kappa^2}{\kappa^{2}(\kappa^{2}-\tilde\kappa^2)} \frac{\partial}{\partial z}\left( \frac{\ex{- \kappa r}}{r}\right)
\end{equation}
The general solution is 
\begin{align}
c'_A = &\frac{\tilde{\cg}\tilde\kappa^2}{\kappa^{2}-\tilde\kappa^2}  \frac{\partial}{\partial z}\left( \frac{\ex{- \kappa r}}{\kappa^{2}r} + A_1 \frac{\ex{\tilde\kappa r}}{r}+A_2 \frac{\ex{-\tilde\kappa r}}{r}+A_3r^2+\frac{A_4}{r}\right)
\end{align}
For the potential to remain bounded at $r\to \infty$, one imposes $A_1=A_3=0$. Moreover, for $\nabla^2 c'_A$ to remain finite at $r=0$, one sets $A_2=-1/\tilde\kappa^2$. For $c'_A$ to remain finite at $r=0$, one needs $A_4=1/\tilde\kappa^2 - 1/\kappa^2$. We then get
\begin{equation}
c'_A = \frac{\tilde{\cg}\tilde\kappa^2}{\kappa^{2}-\tilde \kappa^2}  \frac{\partial}{\partial z}\left(  \frac{1-\ex{-\tilde\kappa r}}{\tilde\kappa^2 r} -    \frac{1-\ex{-\kappa r}}{ \kappa^2 r}     \right).
\end{equation}
This leads to Eq. \eqref{phi1_exp}.


%

\end{document}